\documentstyle[12pt]{article}

\def\a{{\bf a}}

\def\v{{\bf v}}
\def\x{{\bf x}}

\def\j{{\bf j}}
\def\A{{\bf A}}
\def\B{{\bf B}}
\def\E{{\bf E}}
\def\J{{\bf J}}
\def\R{{\bf R}}
\def\V{{\bf V}}
\def\g{{\bf g}}

\def\<{\langle}
\def\>{\rangle}
\def\del{\nabla}

\def\d{\partial}

\def\u{{\bf u}}
\def\U{\,{{\nabla \,\rho} \over \rho}\,}
\def\W{\,{{\J^{\mathrm gi}} \over \rho}\,}
\normalbaselines 
\begin{document} 
\begin{center}
\vspace*{1.0cm}

{\LARGE{\bf Perspectives on Nonlinearity \\
in Quantum Mechanics}} 

\vskip 1.5cm

{\large {\bf Gerald A. Goldin}} 

\vskip 0.5 cm 

Departments of Mathematics and Physics \\ 
Rutgers University \\ 
SERC Bldg. Rm. 238, Busch Campus \\
118 Frelinghuysen Road \\ 
Piscataway, NJ 08854 USA \\
\end{center}

\begin{flushright}
{\it gagoldin@dimacs.rutgers.edu\/}
\end{flushright}

\noindent
{\em It is with great pleasure that I dedicate this contribution
to my friend and collaborator, Prof. Dr. Heinz-Dietrich Doebner, on the
special occasion of his retirement from the Arnold Sommerfeld Institute
for Mathematical Physics.}

\vspace{0.6 cm}

\begin{abstract}

\bigskip
\noindent
Earlier H.-D. Doebner and I proposed a family of
nonlinear time-evolution equations for quantum mechanics associated
with certain unitary representations of the group of diffeomorphisms
of physical space. Such nonlinear Schr\"odinger equations may
describe irreversible, dissipative quantum systems. We subsequently
introduced the group of nonlinear gauge
transformations necessary to understand the resulting
quantum theory, deriving and interpreting gauge-invariant
parameters that characterize (at least partially) the physical content.
Here I first review these and related results, including the coupled
nonlinear Schr\"odinger-Maxwell theory, for which I
also introduce the gauge-invariant (hydrodynamical)
equations of motion. Then I propose a further,
radical generalization. An enlarged group  
${\cal G}$ of nonlinear transformations,
modeled on the general linear group $GL(2,\R)$, leads to a
beautiful, apparently unremarked symmetry
between the wave function's phase and the logarithm of its
amplitude. The equations Doebner and I proposed are
embedded in a wider, natural family of nonlinear
time-evolution equations, invariant
(as a family) under ${\cal G}$. Furthermore there
exist ${\cal G}$-invariant
quantities that reduce to the usual
expressions for probability density and flux
for linearizable quantum
theories in a particular gauge.
Thus ${\cal G}$ may be interpreted as generalizing further
our notion of nonlinear gauge transformation.

\end{abstract}

\vspace{1 cm} 

\section{Families of Nonlinear Schr\"odinger Equations}

\noindent
About nine years ago, H.-D. Doebner and I introduced
a certain family of nonlinear Schr\"odinger equations.
We were led to these equations not by any prior inclination
to study nonlinear quantum mechanics, but by our desire
to interpret quantum-mechanically a class of
representations of an infinite-dimensional,
nonrelativistic current algebra, and the corresponding
group \cite{DG1992,G1992,DG1994b}. We proposed these equations
as candidates for describing
quantum systems with dissipation.

To review the development briefly, we sought self-adjoint
representations of the infinite-dimensional Lie algebra of
densities and currents, given at arbitrary time $t$ by
\[
[\,\rho_{op} (f_{1}),\, \rho_{op} (f_{2})] \, = \, 0\,,\;\;\;
[\,\rho_{op} (f), \, J_{op}(\g)] \,=\,
i\hbar \rho_{op} (\g \cdot \del f)\,,
\label{currentalg1}
\]
\begin{equation}
[J_{op}(\g_{1}),\,J_{op}(\g_{2})] \,=\,
-i\hbar J_{op}([\g_{1},\, \g_{2}])\,,
\label{currentalg2}
\end{equation}
where the $f$'s are real-valued
$C^{\,\infty}$ functions on the physical space $\R^{\,n}$,
the $\g$'s are $C^{\,\infty}$ vector fields on  $\R^{\,n}$, and
$\,[\,\g_{1},\g_{2}\,] \,=\,
\g_{1}\cdot\nabla\g_{2} - \g_{2}\cdot\nabla\g_{1}\,$
is the usual Lie bracket \cite{DS1968,GS1969,G1971,DT1980}.
The $N$-particle Bose or Fermi representations
of (\ref{currentalg2}) may be written
\begin{eqnarray}
\rho_{op}^{\,N}(f)\,\psi^{(s,a)} 
(\x_1, \dots \x_N) \,\,=\,\,
m\sum_{j=1}^{N} f(\x_j)\psi^{(s,a)}
(\x_1, \dots \x_N),\quad\nonumber\\
J_{op}^{\,N}(\g)\,\psi^{(s,a)} 
(\x_1, \dots \x_N) \,\,=\,\, 
\frac{\hbar}{2i}
\,\sum_{j=1}^{N} \,\{\,\g(\x_j)\cdot
\nabla_j \psi^{(s,a)}(\x_1, \dots \x_N)
\quad\quad\quad \nonumber\\
\quad\quad\quad\,+\,\,\nabla_j\cdot
[\,\g(\x_j)\psi^{(s,a)}
(\x_1, \dots \x_N)\,]\,\}\,,
\label{Nparticle}
\end{eqnarray}
where the $\,\psi^{(s,a)}\,$ are (respectively)
symmetric or antisymmetric
square-integrable functions of the $N$
particle coordinate variables.
There exists a family of related but
unitarily inequivalent representations
of (\ref{currentalg2}),
parameterized by the real number $D\/$,
leading to physically distinct
quantizations \cite{GMS1981b,ADT1983}:
\begin{equation}
J_{op}^{\,N,D}(\g) \,=\, J_{op}^{\,N}(\g) \,+\,
D\,\rho_{op}^{\,N}(\nabla \cdot \g).
\label{Dreps}
\end{equation}
Here $D$ is a constant
with the dimensions of a diffusion coefficient.
Even in the case of one-particle quantum mechanics,
interpreting these representations
posed a challenge.

In the usual
notation for operator-valued distributions, write
(suppressing the superscripts)
$\,\rho_{op}(f) = \int_X \rho_{op}(\x)f(\x)d\x\,$ and
$\,J_{op}(\g) = \int_X \J_{op}(\x)\cdot\g(\x)d\x$.
Then, for a single particle at time $t$, take the
expectation values
$\,m\,\rho\,(\x,t) \,=\,
\<\psi_t|\,\rho_{op}(\x)\,|\psi_t\>\,$
and $m\,\j\,(\x,t) \,=\,
\<\psi_t|\,\J_{op}(\x)\,|\psi_t\>$.
When $\,D = 0\,$ the usual expressions are recovered
for the probability density
and flux in the Schr\"odinger representation:
\begin{equation}
\rho \,=\, \overline{\psi} \psi\,, \quad
{\j} \,=\, \frac{\hbar}{2mi}\,[\,\overline{\psi} \nabla \psi
- (\nabla \overline{\psi}) \psi\,]\,.
\label{rhoj}
\end{equation}
For arbitrary $D$, one obtains instead
$\,\j^{\,D} \,=\, \j \,-\,D\, \nabla (\overline{\psi} \psi)$.
Imposing the equation of continuity
$\,\partial_{\,t}\,\rho \,=\, - \nabla \cdot \j^{\,D}\,$
then gives,
as a kinematical
constraint on the time-evolution of $\,\psi\,$,
a Fokker-Planck type of equation:
$\,\partial_{\,t}\,\rho \,=\, - \nabla \cdot \j \,+\,
D \, \nabla^2 \rho\,$.

No linear time-evolution equation for
$\psi$ obeys this constraint.
Rather we derived an interesting family of
nonlinear Schr\"odinger equations,
with the purely imaginary functional
$\,i \hbar (D/2)\,\nabla^2 \rho /\rho\,$ multiplying
$\psi$ on the right-hand side. That is, this
particular form of nonlinearity
was forced on us by the current algebra representation.
And without linearity as an axiom, we also could not
eliminate {\it a priori\/} the possibility of
additional, real nonlinear functionals
multiplying $\psi$. Doebner and I restricted
these to homogeneous rational expressions with no
more than two derivatives in the numerator.
Defining (for convenience) $\,\hat{\j} \,=\,
(m / \hbar)\,\j\,=\, (1/2i)\,[\,\overline{\psi} \nabla \psi
- (\nabla \overline{\psi}) \psi\,]\,$, we introduced
the real, homogeneous
functionals $\,R_1[\psi], \dots, R_5[\psi]\,$
given by
\begin{equation}
R_1 \,=\,
{\nabla \cdot {\hat{\j}} \over \rho}\,, \quad
R_2 \,=\,
\,{\nabla^{\,2} \rho \over \rho}\,, \quad
R_3 \,=\,
{\hat{{\j}}^{\,2}\over \rho^2}\,, \quad
R_4 \,=\,
{\hat{\j} \cdot \nabla\rho \over \rho^2}\,,\quad
R_5 \,=\,
{(\nabla \rho)^2 \over \rho^2}\,.
\label{Rjdefs}
\end{equation}
The family of nonlinear Schr\"odinger equations
became then:
\begin{equation}
i\hbar \frac{\d \psi}{\d t} \,=\,
H_{0}\,\psi
\,+ \frac{\,i\,}{2} \hbar D\,R_2[\psi]\, \psi \,+\,
\hbar D^{\,\prime} \sum_{j=1}^5\, c_j\, R_j[\psi]\,\psi\,,
\label{nlse}
\end{equation}
where $D^{\,\prime}$ is another
diffusion coefficient, the $c_j$ are
real and dimensionless, and
\begin{equation}
H_0\,{\psi} \,=\,
{1 \over 2m}\,
[-i\hbar\nabla - (e/c) \A(\x,t)]^2\,\psi \,+\,
[\,V\,+\,\,e\,\Phi(\x,t)]\,\psi\,.
\label{lse}
\end{equation}
Below we shall see how an important subclass of
(\ref{nlse}), and certain more
general nonlinear Schr\"odinger equations, can be
obtained from the linear Schro`\"odinger equation
via nonlinear gauge transformations.
Eq. (\ref{nlse}) contains as special cases
a remarkable variety of nonlinear modifications
of quantum mechanics proposed independently
by other researchers
\cite{Kib1978,GP1982,SCH1983,StYS1989,Sab1990,MSt1991,AuSab1994},
though without our fundamental motivation for the
nonlinearity and typically without the above
local, pure imaginary
nonlinear functional multiplying $\,\psi$.

Using the expansion $\,\del^2\,\psi/\psi \,=\, i R_1[\psi]
\,+\,(1/2)\,R_2[\psi]\,-\, R_3[\psi]
\,-\, (1/4)\,R_5[\psi]\,$, let us rewrite this family
of equations as in Ref. \cite{G1997}, with
some additional terms:
\[
i\,{{\dot{\psi}} \over {\psi}}\,\,\,\,=\,\,\,\,
i\left[\,\,\sum_{j=1}^2 \nu_j R_j[\psi] \,+\,
{{\nabla \cdot ({\cal A}(\x,t)\rho)}\over \rho}\,\right] \,+\,
\]
\begin{equation}
\left[\,\,\,\sum_{j=1}^5 \mu_j R_j[\psi] \,+\,
\, U(\x,t)\,+\,
{{\nabla \cdot ({\cal A}_1(\x,t)\rho)}\over \rho} \,+\,
{{{\cal A}_2(\x,t) \cdot \hat{\j}}\over \rho} \,+\, 
\alpha_1 \,\ln \rho \,+\, \alpha_2\,S\,
\right].
\label{nlsegen}
\end{equation}
Here $S$ is the phase of $\psi$,
$U$ is a (sufficiently smooth) external, real-valued,
time-dependent scalar function;
and ${\cal A},\,{\cal A}_1,$ and ${\cal A}_2$ are
distinct (sufficiently smooth)
external, real-valued, time-dependent
vector fields.
Eq. (\ref{nlse}) is obtained from
Eq. (\ref{nlsegen}) with the following
substitutions:
\[
\nu_1 = - \frac{\hbar}{2m}\,,\quad
\nu_2 = \frac{1}{2} D\,,\quad
{\cal A} = {e \over 2mc} \A\,,
\]
\[
\mu_1 = D^{\,\prime}c_1\,,\quad
\mu_2 = - \frac{\hbar}{4m} +  D^{\,\prime}c_2\,,
\quad
\mu_3 =
\frac{\hbar}{2m} +  D^{\,\prime}c_3\,,
\quad
\mu_4 =  D^{\,\prime}c_4\,,
\quad
\mu_5 = \frac{\hbar}{8m} +
D^{\,\prime}c_5\,,
\]

\smallskip
\noindent
\[
U(\x, t) = \frac{1}{\hbar}\,[\,V(\x,t)\,+\,
e\,\Phi\,]\,+\,\frac{e^2}{2m\hbar c^2}\,\A^2,\quad
{\cal A}_1 = 0\,,\quad
{\cal A}_2 = -{e \over mc} \A\,,
\]

\medskip
\noindent
\begin{equation}
\alpha_1 \,=\, \alpha_2 \,=\, 0.
\label{numuvals}
\end{equation}
The coefficients $\nu_j\,\,(j = 1,2),$
$\mu_j\,\,(j = 1, \dots\,, 5),$
and $\alpha_j\,\,(j = 1,2)$
are taken to be continuously
differentiable, real-valued functions of $\,t\,$.
The motivation for this expansion, the
reason behind the introduction of terms with $\,\alpha_1\,$,
$\,\alpha_2\,$, and $\,{\cal A}_1\,\not=\,0$,
and the reason for permitting the coefficients to be
time-dependent, all stem from the discussion
of nonlinear gauge transformations
in the next section.

Finally, let us introduce here a further,
natural generalization of Eq. (\ref{nlsegen}).
Let us insert  into the imaginary part
of the right-hand side the terms
$\,\nu_3 R_3,\,\,\nu_4 R_4$, and $\,\nu_5 R_5$, as well
as new external scalar and vector fields, 
to achieve full symmetry between the real
and imaginary parts \cite{G1999}. Thus
we have, in effect, allowed for {\em complexification\/} of all
the coefficients and external fields. The equation becomes:
\[
i\,{{\dot{\psi}} \over {\psi}}\,=\,
i\left[\,\,\sum_{j=1}^5 \nu_j R_j[\psi] \,+\,
\, {\cal T}(\x,t) \,+\,
{{\nabla \cdot ({\cal A}(\x,t)\rho)}\over \rho} \,+\,
{{{\cal D}(\x,t) \cdot \hat{\j}}\over \rho} \,+\,
\delta_1 \,\ln \rho \,+\, \delta_2\,S\,\right]\,+\,
\]
\begin{equation}
\left[\,\,\,\sum_{j=1}^5 \mu_j R_j[\psi] \,+\,
\, U(\x,t) \,+\,
{{\nabla \cdot ({\cal A}_1(\x,t)\rho)}\over \rho} \,+\,
{{{\cal A}_2(\x,t) \cdot \hat{\j}}\over \rho} \,+\, 
\alpha_1 \,\ln \rho \,+\, \alpha_2\,S\,
\right],
\label{nlsegen2}
\end{equation}
where $\,{\cal T}$ is a new external scalar field, and
$\,{\cal D}$ a new external vector field. Note that
the heat equation and other interesting equations of
mathematical physics fall within this family.
Some equations with
soliton-like solutions are also included \cite{CaDoMi1998}.
But the equation
of continuity relating $\,\rho\,$ and
$\,\j^{\,D}\,$ no longer holds.
Evidently when $\nu_3 = \nu_4 = \nu_5
= \delta_1 = \delta_2 = 0$,
${\cal T} = 0$, and ${\cal D} = 0$,
we recover Eq. (\ref{nlsegen}).
When the remaining values are
as in Eq. (\ref{numuvals}) with $D = D^{\,\prime} = 0$,
we are back with the linear Schr\"odinger equation.

We shall see
that the generalization of Eq. (\ref{nlsegen})
to Eq. (\ref{nlsegen2})
follows from a further,
natural extension of the notion of nonlinear gauge
transformation.

% end of Section 1 

\section{Time-Dependent Nonlinear Gauge\\
Transformations} 

\noindent
Let us write $\,\psi \,=\,R\,\exp\,[\,iS\,]\,$,
where the amplitude $R$ and the
phase $S$ are real. Then
$\,\rho \,=\, R^{\,2}\,$
and $\,\j \,=\, (\hbar/m)\,R^2\, \nabla S\,$.
While $R$ is gauge invariant, $S$ is not:
under the usual, unitary gauge transformations of
quantum mechanics, $\,R^{\,\prime} = R\,$ but
$\,S^{\,\prime} \,=\, S \,+\, \theta(\x,t)\,$.
Then $\,\rho^{\,\prime} \,=\, \rho\,$,
while $\,\j^{\,\prime} \,=\, \j \,+\,
(\hbar/m)\,R^{\,2}\,\del\theta\,$.

If we begin with the linear Schr\"odinger equation
in the absence of a vector potential, i.e.,
$\,i\hbar \partial_{\,t}\, \psi \,=\,
-\,(\hbar^2/2m)\,\nabla^{\,2}\,\psi \,+\,V\,\psi\,$,
then the transformed
wave function $\,\psi^{\,\prime} \,=\,R^{\,\prime}
\exp\,[\,iS^{\,\prime}\,]\,$ satisfies
$\,\,i\hbar \partial_{\,t}\, \psi^{\,\prime}
\,=\, (\hbar^2/2m)\,[-i\nabla \,-\,
{\mathrm grad}\,\theta]^2\,
\psi^{\,\prime}
\,+\, [V \,-\, \hbar \dot{\theta}]\,\psi^{\,\prime}$.
This observation can actually motivate
introduction of the
external electromagnetic gauge
potentials $\,\A\,$ and $\,\Phi\,$, and
the ``minimally coupled''
Schr\"odinger equation whose Hamiltonian is given by
Eq. (\ref{lse}). When we begin with
(\ref{lse}), we have that $\,\psi^{\,\prime}$ satisfies
the transformed equation obtained
by substituting the gauge-transformed potentials:
$\,\A^{\,\prime} \,=\,
\A \,+\, (\hbar c/e)\,{\mathrm grad}\,\theta\,$
and $\,\Phi^{\,\prime} \,=\,
\Phi \,-\, (\hbar / e) \, \dot{\theta}\,.$
A gauge-invariant current can now be
written $\,\J^{\rm{gi}} \,=\, \j - (e/mc)\,\rho\, \A\,$,
with $\,\partial_{\,t}\,\rho \,=\, -\,\nabla \cdot
\J^{\rm{gi}}\,$.
The physical fields $\,\B \,=\, \nabla \times \A\,$
and $\,\E \,=\, -\,\nabla\Phi - (1/c) \,\partial_{\,t}\,{\A}\,$
are likewise gauge invariant. All this is 
elementary, and standard. It sets the pattern
for consideration of nonlinear gauge transformations
for nonlinear Schr\"odinger equations.

In the latter context we
(necessarily) abandon
the usual, tacit assumption that gauge
transformations act linearly
and unitarily. Doebner and I introduced
a group of nonlinear transformations
leaving our class of equations invariant as a family
\cite{DG1996,G1997},
\begin{equation}
R^{\,\prime} \,=\, R\,,\quad
S^{\,\prime} \,=\, \Lambda S + \gamma \ln R + \theta\,,
\label{RSlaw}
\end{equation}
where in general $\,\gamma\,$ and $\,\Lambda\,$
are continuously
differentiable, real-valued
functions of $\,t\,$, $\,\Lambda \,\not=\, 0$,
and $\,\theta\,$ is a continuously differentiable, real-valued
function of $\,\x\,$ and $\,t$.
Then $\,(\Lambda_1,\,\gamma_1,\,\theta_1)\,
\,(\Lambda_2,\,\gamma_2,\,\theta_2)\,=\,
(\Lambda_1\Lambda_2,\,
\gamma_1 \,+\, \Lambda_1\gamma_2,\,
\theta_1 \,+\, \Lambda_1\theta_2)$.
The original justification for taking these to be
gauge transformations was
the argument, put forth by many theorists, that
any physical quantum-mechanical
measurement could be reduced to
a sequence of positional measurements
at different times; with the system subjected
to external force fields between
measurements \cite{FH1965,Miel1974}.
Under Eq. (\ref{RSlaw}),
\[
\rho^{\,\prime} \, = \,\overline{\psi^{\,\prime}}\,\psi^{\,\prime} \,=
\, \rho\,,
\]
\begin{equation}
\hat{\j}^{\,\prime}
\, = \,
\frac{1}{2i}\,[\,\overline{\psi^{\,\prime}}\,\nabla \psi^{\,\prime}
- (\nabla \overline{\psi^{\,\prime}}\,)\,\psi^{\,\prime}\,]\,
\, = \,
\Lambda\,\hat{\j}\, + \,\frac{\gamma}{2} \nabla \rho \,+\, \rho \nabla \theta\,.
\label{rhojtransform}
\end{equation}
Keeping the interpretation of
$\,\rho \,=\, |\psi|^2\,$
as the positional probability density,
and writing invariant force fields
in terms of the external potentials,
the outcomes of all measurements
do remain invariant. Eq. (\ref{RSlaw}) also
has other nice properties: it is strictly
local, and it respects a certain separation
condition for (many-particle)
product wave functions
\cite{GSv1994,DGN1999}.
If $\,\psi\,$ obeys
a Schr\"odinger equation
of the type in Eq. (\ref{nlsegen}), then
$\psi^{\,\prime}$ transformed by
(\ref{RSlaw}) obeys another equation in the family,
with transformed coefficients
and external fields. The
coefficients are given by:
\[
\nu_1^{\,\prime} = \frac{\nu_1}{\Lambda}\,,\quad
\nu_2^{\,\prime} = -\frac{\gamma}{2\Lambda}\nu_1 +\nu_2\,,
\]
\[
\mu_1^{\,\prime} = -\frac{\gamma}{\Lambda}\nu_1 + \mu_1\,,\quad
\mu_2^{\,\prime} = \frac{\gamma^2}{2\Lambda}\nu_1-\gamma \nu_2
- \frac{\gamma}{2}\mu_1+\Lambda \mu_2\,,
\]

\smallskip
\noindent
\[
\mu_3^{\,\prime} =
\frac{\mu_3}{\Lambda}\,,\quad
\mu_4^{\,\prime}= -\frac{\gamma}{\Lambda}\mu_3 + \mu_4\,,\quad
\mu_5^{\,\prime} = \frac{\gamma^2}{4\Lambda}\mu_3
- \frac{\gamma}{2}\mu_4
+ \Lambda\mu_5\,,
\]

\smallskip
\noindent
\begin{equation}
\alpha_1^{\,\prime} \,=\,
\Lambda \alpha_1 \,-\, {\gamma \over 2}\,\alpha_2\,+\,
{1 \over 2}\,\left({\dot{\Lambda} \over \Lambda}\gamma
\,-\, \dot{\gamma}\right),\quad
\alpha_2^{\,\prime} \,=\,
\alpha_2 \,-\,{\dot{\Lambda} \over \Lambda}\,,
\label{primescorr}
\end{equation}

\medskip
\noindent
while the transformed vector and scalar fields are

\bigskip
\noindent
\[
{\cal A}^{\,\prime} \,=\, {\cal A}
\,-\, \frac{\nu_1}{\Lambda}\,\nabla \theta\,,
\]

\smallskip
\noindent
\[
{\cal A}_1^{\,\prime} \,=\,
\Lambda \, {\cal A}_1\, -\, \gamma \, {\cal A} \,-\,
{\gamma \over 2}\,{\cal A}_2 \,+\,
\left(\,{\gamma \over \Lambda}\,\nu_1 - \mu_1
+ {\gamma \over \Lambda}\, \mu_3
- \mu_4\right)\,\nabla \theta\,,
\]

\medskip
\noindent
\[
 {\cal A}_2^{\,\prime} \,=\,
{\cal A}_2 - \frac{2\mu_3}{\Lambda}\,\nabla \theta\,,
\]
\[
U^{\,\prime} \,=\,\Lambda\,U \,-\, \dot{\theta}
\,+\,\left({\dot{\Lambda} \over \Lambda} - \alpha_2\right)\theta
\,+\,{{\mu_3} \over \Lambda}\,[\,\nabla \theta\,]^{\,2} \,+\,
\quad\quad\quad\quad\quad\quad
\]

\smallskip
\noindent
\begin{equation}
\quad\quad\quad\quad
\left(\mu_4 - \mu_3\,{\gamma \over \Lambda}\,\right)
\,\nabla^2\theta
\,+\,{\gamma \over 2}\,\nabla\cdot{\cal A}_2
\,-\,{\cal A}_2 \cdot \nabla \theta.
\label{potprimes}
\end{equation}

\medskip
\noindent
Regarding Eqs. (\ref{primescorr}),
note how the time-dependence of $\,\gamma\,$
and $\,\Lambda\,$ in Eq. (\ref{RSlaw})
{\it requires\/} that the
$\,\nu_j,\,\mu_j\,$, and
$\,\alpha_j\,$ in Eq. (\ref{nlsegen}) be
time-dependent, and that the $\,\alpha_j\,$
be allowed nonzero values. The terms with
$\,\alpha_1\,$ and $\,\alpha_2\,$ were,
respectively, first introduced by Bialynicki-Birula
and Micielski \cite{BM1976} and by
Kostin \cite{Kos1972}. Likewise, we see in
(\ref{potprimes}) how the
$\,{\cal A}_1\,$ and $\,{\cal A}_2\,$ terms in
Eq. (\ref{nlsegen}) are needed. Nonlinear
Schr\"odinger equations with arbitrary values
of $\,{\cal A}_2\,$ were considered by
Haag and Bannier \cite{HB1978},
while as far as I know the field $\,{\cal A}_1\,$
was first considered in Ref. \cite{G1997}.
An important subclass of Eq. (\ref{nlsegen}) is linearizable
by means of nonlinear gauge transformations; for this
subclass, the physics is unchanged from ordinary
quantum mechanics.

The coefficients, the external fields, and many of the
nonlinear functionals in Eq. (\ref{nlsegen})
are not gauge invariant. But we do have a
current $\,\J^{{\mathrm gi}}\,,$
invariant under nonlinear gauge
transformations, that
enters the continuity equation
$\,\dot{\rho} = - \nabla \cdot \J^{{\mathrm gi}}$,
given by
\begin{equation}
\,\J^{{\mathrm gi}}\,=\,
-\,2\nu_1\,\hat{\j} \,-\,2 \nu_2 \nabla \rho
\,-\, 2 \rho {\cal A}\,.
\label{Jgi}
\end{equation}
This reduces, of course, to the usual gauge-invariant
current in the linear case
\cite{G1997}. Now, the existence of
$\,\J^{{\mathrm gi}}\,$ means
that our earlier assumption about all measurements
being reducible to a succession of
positional measurements
is unnecessarily restrictive. It is sufficient
that all measurements be expressible in terms
of gauge-invariant quantities; and we have
available for this the density $\,\rho\,$, the
current $\,\J^{{\mathrm gi}}\,$, and gauge-invariant
force fields (see below).

Doebner and I also introduced gauge-invariant parameters:
\[
\tau_1 = \nu_2 - \frac{1}{2}\mu_1\,,\quad
\tau_2 = \nu_1\mu_2 -\nu_2\mu_1\,,\quad
\tau_3 = \frac{\mu_3}{\nu_1}\,,\quad
\tau_4 = \mu_4 - \mu_1 \frac{\mu_3}{\nu_1}\,,
\]
\[
\tau_5 = \nu_1\mu_5 - \nu_2\mu_4 +
\nu_2^{\,2}\,\frac{\mu_3}{\nu_1}\,,\quad
\]
\begin{equation}
\beta_1 \,=\, \nu_1 \, \alpha_1 \,-\, \nu_2 \,\alpha_2 \,+\,
\nu_2\, {\dot{\nu}_1 \over \nu_1} - \dot{\nu}_2\,,\quad
\,\beta_2 \,=\, \alpha_2 \,-\,  {\dot{\nu}_1 \over \nu_1}\,.
\label{invariants}
\end{equation}
Some discussion of the physics behind these parameters
may found in Ref. \cite{DG1996}; in particular,
$\tau_1 \not= 0$, $\tau_4 \not= 0$, or $\beta_2 \not= 0$
violates time-reversal invariance; $\tau_3 \not= -1$ or
$\tau_4 \not= 0$ breaks Galileian invariance; and
in all these cases $\tau_2$ corresponds to the
{\it observed\/} value of $\,\hbar^2/8m^2$
(no longer can we identify the gauge-dependent
quantity $\,-\nu_1\,$ with the gauge-independent,
observable constant $\,\hbar/2m\,$).
Thus the classical limit can be taken in
a gauge-invariant manner by letting $\tau_2 \to 0$.

Let me also remark here that the gauge-invariant
parameter $\,\beta_2\,$ is naturally interpreted as
a coefficient of friction, as it contributes (see below)
a term $\,- \beta_2\,(\J^{\mathrm gi}/\rho)\,$ to
the expression for $\,\partial_t\,(\J^{\mathrm gi}/\rho)$.

Continuing the discussion in Ref. \cite{G1997}
we have also gauge-invariant fields. Set
\begin{equation}
\hat{U} \,=\,-\,\nu_1\,U \,-\,
\tau_3\,{\cal A}^{\,2}
\,-\, (\tau_4 - 2\tau_1\tau_3)\,\nabla \cdot {\cal A}
\,+\,{\cal A}\cdot{\cal A}_2
\,-\, \nu_2\,\nabla \cdot {\cal A}_2\,,
\label{Uhatcorrected}
\end{equation}
so that under nonlinear gauge transformation,
\begin{equation}
\hat{U}^{\,\prime}\,=\,\hat{U} \,+\,
{\nu_1 \over \Lambda}\,\dot{\theta}\,+\,
{\nu_1 \over \Lambda}\,\alpha_2\,\theta\,-\,
\nu_1\,{\dot{\Lambda} \over \Lambda^2}\,\theta\,.
\label{Uprimenew}
\end{equation}
Eq. (\ref{Uhatcorrected}) corrects algebraic
errors in Ref. \cite{G1997}.
The field $\,\hat{U}\,$ is easily reduced to
$\,(1/2m)\,(V \,+\, e\,\Phi)\,$
for the linear Schr\"odinger equation.
We have the new gauge-invariant vector fields,

\medskip
\noindent
\[
{\cal A}_1^{\,\,gi} \,=\,\nu_1 {\,\cal A}_1 \,+\,
\left(\frac{2\nu_2\mu_3}{\nu_1} - \mu_1 - \mu_4\right)\,{\cal A} \,-\,
\nu_2\,{\cal A}_2\,,
\]

\medskip
\noindent
\begin{equation}
{\cal A}_2^{\,\,gi} \,=\,
{\nu_1 \over 2\mu_3}\,{\cal A}_2 - {\cal A}\,,
\label{gaugeinvA}
\end{equation}
as well as magnetic and (generalized) electric
plus other potential force fields,

\bigskip
\noindent
\[
{\cal B} = \nabla \times {\cal A}\,= \,{e \over 2mc}\,\B,
\]
\begin{equation}
{\cal E} \,=\, - \,\nabla\hat{U} -
\frac{\partial {\cal A}}{\partial t}\,-\,
\beta_2\,{\cal A}\,=\, - {1 \over 2m} \nabla V
\,+\, {e \over 2m}\,\E.
\label{Enew}
\end{equation}
Thus $\,\hat{U}\,=\,(1/2m)(V + e\,\Phi)\,$
in general, and
$\,\E\,=\,-\nabla \Phi \,-\, (1/c)\,\partial_t \A \,-\,
(\beta_2/c)\,\A$.
Notice the extra term associated with
Kostin's nonlinearity; without it,
$\,{\cal E}\,$ is not
gauge invariant. This
leads in turn to an interesting modification
of one of Maxwell's equations:
\begin{equation}
\,\nabla \times {\E}\,=\,
-\,{1 \over c}\,{\partial\,{\B} \over \partial t} \,-\,
{\beta_2 \over c}\,{\B}\,.
\label{Maxnew}
\end{equation}
%

% end of Section 2 

\section{Gauge-Invariant Equations of Motion}

Using the (hydrodynamical) variables
$\,\rho\,$ and $\,\V = \J^{\mathrm gi}/\rho\,$,
it is straightforward to write down
in manifestly gauge-invariant form
the equations of motion corresponding
to Eq. (\ref{nlsegen}).
We have in all cases the useful relation
$\,\nabla \times \V\,=\,-2{\cal B}\,=\,(e/mc)\B$,
and the continuity equation
$\,\partial_t \,\rho \,=\, - \nabla \cdot \J^{\mathrm gi}$.
In addition,

\[
{\partial \over {\partial t}}
\left(\,{\J^{\mathrm gi} \over \rho}\,\right)
\,=\, \nabla \left[ \,2\tau_1\,\nabla \cdot \left(\W \right)
\,+\, 2 \tau_2\,{{\nabla^2\,\rho} \over \rho}
\,+\,{1 \over 2}\,\tau_3\left(\W \right)^2\,\right]
\]

\[
\,+\,\nabla \left[ \,(\,2\tau_1\,[1 + \tau_3]\,- \tau_4\,)
\left(\W \right)\,\cdot \U
\,+\,2\tau_5 {(\nabla \rho)^2 \over {\rho^{\,2}}}\,\,\right]
\]

\[
\,+\,\,\nabla \left[\,\,2\,
{\nabla \cdot {({\cal A}_1^{\,\,gi}\rho)} \over \rho}
\,-\,2\tau_3\,{\cal A}_2^{\,\,gi}\cdot \left(\W \right)
\,+\,2 \beta_1 \,\ln \rho\,\,\right]
\]

\begin{equation}
\,-\, \beta_2\,\left(\W \right)
\,-\,{1 \over m} \nabla V
\,+\, {e \over m}\,\E.
\label{gimotion}
\end{equation}

\bigskip
\noindent
Now we have the expected values of position,
velocity, and acceleration:
\[
<\x> \,=\, \int \x\,\rho\,(\x)\,d\x\,,
\]

\begin{equation}
<\v> \,=\, {{\partial <\x>} \over {\partial t}}
\,=\, \int \rho\,\left(\W \right)\,d\x\,,
\label{forcelaws}
\end{equation}

\[
<\a> \,=\, {{\partial <\v>} \over {\partial t}}
\,=\, \int \rho \,\left[
\,{1 \over 2}\,\nabla \left(\W \right)^2
\,+\,\left(\W \right)\,\times \,{e \over {mc}}\,\B\,
\,+\,{\partial \over {\partial t}}
\left(\W \right)\,\,\right]\,d\x\,.
\]

\smallskip
\noindent
Note that in Eqs. (\ref{gimotion})-(\ref{forcelaws}),
the force laws governing interaction with the external
electric and magnetic fields are unchanged from linear
quantum mechanics.

\section{The Enlarged Gauge Group}

\noindent
To this point, the amplitude $R$ and
the phase $S$ have a fundamentally
different status, both in linear quantum mechanics
and in our nonlinear variations:
$R$ is gauge invariant, and physically
observable; while $S$ is not. This asymmetry seems
more and more puzzling as one comes to appreciate
the flexibility of description offered by nonlinear
quantum time-evolutions, allowing for instance
linear quantum
mechanics to be written in a nonlinear gauge.
Why should we be required to combine the
{\it gauge\/} field $\,S\,$ with the {\it physical\/}
field $\,R\,$ into a
single complex-valued function $\,\psi\,$,
and then through the Schr\"odinger equation
couple both $\,R\,$ and $\,S\,$
to the gauge potentials?
Why not instead try to couple gauge-dependent
quantitites to each other, and correspondingly,
physical fields to each other?

In addition, we remark that
just as the formula (\ref{Jgi}) for the gauge-invariant
current $\J^{\mathrm gi}$ depended on two coefficients
and one external potential in the nonlinear
time-evolution equation (\ref{nlsegen}), there is no
{\it a priori\/} principle that forbids
the formula for the gauge-invariant
{\it probability density\/} from
likewise depending on
coefficients and external potentials
in the time-evolution equation. This is
important as we consider enlarging the
nonlinear gauge group further.

To achieve the desired generalization,
define $\,T\,=\,\ln\,R$, so that $\,\ln\,\psi
\,=\,T \,+\, iS$, and consider the transformations
\begin{equation}
\pmatrix{S^{\,\prime} \cr T^{\,\prime}} \,=\,
\pmatrix{\Lambda & \gamma \cr \lambda & \kappa}
\pmatrix{S \cr T}
+ \pmatrix{\theta \cr \phi}\,,
\label{gnlgt}
\end{equation}
where $\,\Lambda,\,\gamma,\,\lambda,\,$ and
$\,\kappa\,$ depend on $\,t$, and
where $\,\theta\,$ and $\,\phi\,$
depend on $\,\x\,$ and $\,t$. In place of
the condition $\,\Lambda\,\not=\,0$, we
impose that $\,\Delta\,=\,
\kappa \Lambda - \lambda \gamma \not= 0$,
so that (\ref{gnlgt}) is invertible.
This is the transformation group $\,{\cal G}\,$,
modeled on $\,GL(2,\R)\,$,
with which we shall now work;
the earlier gauge group is the subgroup
with $\,\lambda \,\equiv\, 0$,
$\,\kappa \,\equiv\, 1$, and $\phi \,\equiv\,0$.
We thus treat the phase and the
logarithm of the amplitude on an equal footing.
The logarithmic variables $\,T\,$ and $\,S\,$
are, of course, familiar from earlier hydrodynamical
and stochastic versions of
quantum mechanics \cite{N1985,Wa1994};
but they normally are treated quite asymmetrically.

We immediately see that Eq. (\ref{nlsegen})
must be generalized further for it to be
invariant under ${\cal G}$.
This is accomplished by complexifying
the coefficients and external potentials,
to obtain Eq. (\ref{nlsegen2})---a procedure
that is natural, as Eq.
(\ref{gnlgt}) can be obtained by complexifying
$\,\Lambda\,$, $\,\gamma$, and $\,\theta\,$ in the
transformation from $\,\psi\,$ to $\psi^{\,\prime}$.

Since so many terms in our equations involve logarithmic
derivatives, let us continue with the
variables $\,S\,$ and $\,T$. The operation of multiplying
$\,\psi\,$ by a complex scalar is then to
add real constants to $\,S\,$ and to $\,T\,$.
The homogeneous terms in Eq. (\ref{Rjdefs}) become,
$\,R_1\,=\,\del^2\,S \,+\, 2 \del S\cdot\del T\,$,
$\,R_2\,=\,2 \del^2\,T \,+\, 4 (\del T)^2\,$,
$\,R_3\,=\,(\del S)^2\,$,
$\,R_4\,=\,2\del S\cdot\del T\,$, and
$\,R_5\,=\,4(\del T)^2\,$.
We now write the new, general nonlinear Schr\"odinger
equation (\ref{nlsegen2}) as
a pair of coupled partial differential equations
for the extended real-valued functions
$\,S\,$ and $\,T$, which are first order in time
but have general second-order
and quadratic terms:
\begin{eqnarray}
\dot{S} & = & a_1\del^2S \,+\, a_2 \del^2T \,+\, a_3 (\del S)^2
\,+\, a_4 \del S \cdot \del T \,+\, a_5 (\del T)^2 \nonumber \\
&  & \quad \quad \,+\,\, a_6 S \,+\, a_7 T
\,+\, u_0 \,+\, \u_1\cdot\del S
\,+\, \u_2\cdot\del T,\nonumber\\
\dot{T} & = & b_1\del^2S \,+\, b_2 \del^2T \,+\, b_3 (\del S)^2
\,+\, b_4 \del S \cdot \del T \,+\, b_5 (\del T)^2 \nonumber \\
&  & \quad \quad \,+\,\, b_6 S \,+\, b_7 T
\,+\, v_0 \,+\, \v_1\cdot\del S
\,+\, \v_2\cdot\del T.
\label{gcqS}
\end{eqnarray}
The relation between Eq. (\ref{gcqS}) and
and Eq. (\ref{nlsegen2}) is straightforward:
\begin{equation}
\matrix{a_1 = - \mu_1\,, \quad & \quad b_1 = \nu_1\,, \cr
a_2 = - 2\mu_2\,, \quad & \quad b_2 = 2\nu_2\,, \cr
a_3 = - \mu_3\,, \quad & \quad b_3 = \nu_3\,, \cr
a_4 = -2\mu_1 - 2\mu_4\,, \quad & \quad b_4 = 2\nu_1 + 2\nu_4\,, \cr
a_5 = -4\mu_2 - 4\mu_5\,, \quad & \quad b_5 = 4 \nu_2 + 4 \nu_5\,, \cr
a_6 = - \alpha_2\,, \quad & \quad b_6 = \delta_2\,, \cr
a_7 = -2\alpha_1\,, \quad & \quad b_7 = 2\delta_1\,, \cr
u_0 = - U - \nabla \cdot {\cal A}_1\,, \quad & \quad
v_0 = {\cal T} + \nabla \cdot {\cal A}\,, \cr
\u_1 = - {\cal A}_2\,, \quad & \quad \v_1 = {\cal D}\,, \cr
\u_2 = -2{\cal A}_1\,, \quad & \quad \v_2 = 2{\cal A}\,}.
\label{conversion}
\end{equation}
Of course Eq. (\ref{nlsegen}) is embedded
in (\ref{gcqS}), as are many other
interesting equations
of mathematical physics. For reference,
the usual, linear Schr\"odinger equation
(\ref{lse}) corresponds to
\[
a_1 \,=\,0\,,\quad
a_2 \,=\,\frac{\hbar}{2m}\,,\quad
a_3 \,=\,,-\,\frac{\hbar}{2m}\,,\quad
a_4 \,=\,0\,,\quad
a_5 \,=\,\frac{\hbar}{2m}\,,\quad
a_6 \,=\, a_7 \,=\, 0\,,
\]
\[
u_0 \,=\,-\,\frac{1}{\hbar}\,(V + e\Phi)\,
-\,\frac{e^2}{2m\hbar c^2}\,\A^2\,,\quad
\u_1 \,=\,{e \over mc} \A\,,\quad
\u_2 \,=\,0\,,
\]
\[
b_1 \,=\,- \frac{\hbar}{2m}\,,\quad
b_2 \,=\,0\,\quad
b_3 \,=\,0\,\quad
b_4 \,=\,- \frac{\hbar}{m}\,,\quad
b_5 \,=\,0\,,\quad
b_6 \,=\, b_7 \,=\, 0\,,
\]
\begin{equation}
v_0 \,=\,\frac{e}{2mc}\,\nabla \cdot \A,\quad
\v_1 \,=\,0\,,\quad
\v_2 \,=\, {e \over mc} \A\,.
\label{lsevals}
\end{equation}

Now the coefficients $a_j$, $b_j$ obey the following transformation laws
under (\ref{gnlgt}), with the determinant $\Delta = \kappa \Lambda -
\lambda \gamma\,$:

\smallskip
\begin{equation}
\left[\,\matrix{a_1^{\,\prime} \cr a_2^{\,\prime}
\cr b_1^{\,\prime} \cr b_2^{\,\prime}}\,\right] \,=\,
{\Delta}^{-1}
\left[\,\matrix{\kappa\Lambda & -\lambda\Lambda &
\kappa\gamma & -\lambda\gamma
\cr -\gamma\Lambda & \Lambda^2 & -\gamma^2 & \gamma\Lambda
\cr \kappa\lambda & \lambda^2 & \kappa^2 & -\kappa\lambda
\cr -\lambda\gamma & \lambda\Lambda &
-\kappa\gamma & \kappa\Lambda}\,\right]
\left[\,\matrix{a_1 \cr a_2 \cr b_1 \cr b_2}\,\right];
\label{a1a2b1b2prime}
\end{equation}

\medskip
\begin{equation}
\left[\,\matrix{a_3^{\,\prime} \cr a_4^{\,\prime} \cr a_5^{\,\prime}
\cr b_3^{\,\prime} \cr b_4^{\,\prime} \cr b_5^{\,\prime}}\,\right] \,=\,
\Delta^{-2}\,{\Large \cal M}\,
\left[\,\matrix{a_3 \cr a_4 \cr a_5 \cr b_3 \cr b_4 \cr b_5}\,\right],
\quad\rm{where}
\label{a3a4a5b3b4b5prime}
\end{equation}

\smallskip
\[
{\cal M} =
\left[\matrix{\kappa^2\Lambda & -\kappa\lambda\Lambda &
\lambda^2\Lambda & \kappa^2\gamma &
-\kappa\lambda\gamma & \lambda^2\gamma
\cr -2\kappa\gamma\Lambda & \Lambda(\kappa\Lambda + \lambda\gamma) &
-2\lambda\Lambda^2 & -2\kappa\gamma^2 &
\gamma(\kappa\Lambda + \lambda\gamma) & -2\lambda\gamma\Lambda
\cr \gamma^2\Lambda & -\gamma\Lambda^2 &
\Lambda^3 & \gamma^3 & -\gamma^2\Lambda & \gamma\Lambda^2
\cr \kappa^2\lambda & -\kappa\lambda^2 & \lambda^3 &
\kappa^3 & -\kappa^2\lambda & \kappa\lambda^2
\cr -2\kappa\lambda\gamma & \lambda(\kappa\Lambda + \lambda\gamma) &
-2\lambda^2\Lambda & -2\kappa^2\gamma &
\kappa(\kappa\Lambda + \lambda\gamma) & -2\kappa\lambda\Lambda
\cr \lambda\gamma^2 & -\lambda\gamma\Lambda &
-\lambda\Lambda^2 & \kappa\gamma^2 & -\kappa\gamma\Lambda &
\kappa\Lambda^2}\right];
\]

\smallskip
\noindent
and
\begin{equation}
\left[\,\matrix{a_6^{\,\prime} \cr a_7^{\,\prime}
\cr b_6^{\,\prime} \cr b_7^{\,\prime}}\,\right] \,=\,
{\Delta}^{-1}
\left[\,\matrix{\kappa\Lambda & -\lambda\Lambda &
\kappa\gamma & -\lambda\gamma
\cr -\gamma\Lambda & \Lambda^2 & -\gamma^2 & \gamma\Lambda
\cr \kappa\lambda & \lambda^2 & \kappa^2 & -\kappa\lambda
\cr -\lambda\gamma & \lambda\Lambda &
-\kappa\gamma & \kappa\Lambda}\,\right]
\left[\,\matrix{a_6 \cr a_7 \cr b_6 \cr b_7}\,\right]
+\,{\Delta}^{-1}
\left[\,\matrix{\kappa\dot{\Lambda} - \lambda\dot{\gamma}
\cr \Lambda\dot{\gamma} - \gamma\dot{\Lambda}
\cr \kappa\dot{\lambda} - \lambda\dot{\kappa}
\cr \Lambda\dot{\kappa} - \gamma\dot{\lambda}}\,\right].
\label{a6a7b6b7prime}
\end{equation}
The behavior of
the external fields
under generalized gauge
transformation is more complicated.
The transformed vector fields
$\,\u_1^{\,\prime}\,$, $\,\u_2^{\,\prime}\,$,
$\,\v_1^{\,\prime}\,$, and $\,\v_2^{\,\prime}\,$
are linear combinations of the six coefficients
$\,a_3\,$, $\,a_4\,$, $\,a_5\,$, $\,b_3\,$, 
$\,b_4\,$, $\,b_5\,$ and the four vector
fields $\,\u_1\,$,
$\,\u_2\,$, $\,\v_1\,$, and $\,\v_2\,$;
for example, the matrix element of
$\,\u_1^{\,\prime}\,$ by $\,a_3\,$ is
$\,\,\Delta^{-2}\,(-2\kappa^2\Lambda\nabla\theta
\,+\,2\kappa\gamma\Lambda\nabla\phi)$, and its
matrix element by $\,\v_2\,$ is
$\,\,\Delta^{-1}\,(-\lambda\gamma)$.
The transformed scalar fields $\,u_0^{\,\prime}\,$
and $\,v_0^{\,\prime}\,$
are linear combinations of all fourteen
coefficients $\,a_1\,\dots\,a_7\,$
and $\,b_1\,\dots\,b_7\,$, the scalar
fields $\,u_0\,$ and $\,v_0\,$,
and the four vector fields, plus affine
terms that depend on the time-derivatives
of $\,\Lambda\,$ $\,\gamma\,$,
$\,\lambda\,$, $\,\kappa\,$,
$\,\theta\,$, and $\,\phi\,$.
Probably little insight would be
added by reproducing all the equations here.

Now we come to the main point.
The generalization that is proposed will work
(i.e., allow a gauge-invariant theory of measurement)
only if it is possible to write combinations formed
from $S$ and $T$ that are invariant under
Eq. (\ref{gnlgt})---just as the earlier
combinations $\,\rho = \exp [2T]\,$ and
$\,\J^{\mathrm gi} / \rho\,=\,
- 2\nu_1 \nabla S - 4 \nu_2 \nabla T - 2{\cal A}$
are invariant under the smaller group.
Consider for simplicity
only the matrix part of
(\ref{gnlgt}); that is, set $\theta = \phi = 0$;
call the gauge transformation
matrix $\,A\,$. Suppose that
$d_1$, $d_2$ are some
coefficients depending
on the $a_j$ and the $b_j$. Then
$d_1 S + d_2 T$ is invariant under $A$ if and
only if $[d_1\,\,d_2]\,A^{-1} \,=\,
[d_1^{\,\prime}\,\,d_2^{\,\prime}]$. From
(\ref{a3a4a5b3b4b5prime}), we observe
that the choice $d_1 = 2a_3 + b_4$ and
$d_2 = a_4 + 2b_5$ obeys this condition.
Hence $d_1 S + d_2 T$ can serve as one
of the desired invariant combinations.
Next let $L_1 = a_1 S + a_2 T$ and
$L_2 = b_1 S + b_2 T$. Then the
pair $(L_1,\,L_2)$ transforms under $A$
exactly as does the pair $(S, T)$,
whence $d_1 L_1 + d_2 L_2$ is also an invariant.
In fact, any combination
$d_1 (\sigma L_1 + \tau S) + d_2 (\sigma L_2 + \tau T)$,
where $\sigma$ and $\tau$ are fully invariant
combination of the coefficients,
will be invariant; and, of course,
any function of invariants is invariant.
It is straightforward to verify
that $a_1 + b_2 = 2\tau_1$ and
$a_1 b_2 - a_2 b_1 = 2\tau_2$, which
were earlier identified as gauge invariants for
(\ref{RSlaw}), are also invariants
under (\ref{gnlgt}). We shall
interpret $\tau_2 > 0$ as characterizing
the class of Eqs. (\ref{gcqS}) that
pertain to quantum mechanics, with
$\tau_2 \to 0$ defining the classical
limit in a gauge-independent way.

To conclude, the desired invariant
combinations of $S$ and $T$ exist.
There is enough flexibility
to permit a choice that reduces to the
usual formulas in the case of the linear Schr\"odinger
equation. In this way we can construct a
positive definite, gauge-invariant probability
density ${\cal P}^{\,\mathrm gi}$ and
gauge-invariant current
${\bf {\cal J}}^{\mathrm gi}$.
A large subfamily of Eqs. (\ref{nlsegen2})
have solutions for which
${\cal P}^{\,\mathrm gi}$ and
${\bf {\cal J}}^{\mathrm gi}$ obey the desired
continuity equation, so that the total
probability is conserved. And
it is important to stress that a
(smaller) subclass of Eqs. (\ref{nlsegen2})
is equivalent to
ordinary quantum mechanics
by way of generalized nonlinear
gauge transformations, so that we
are assured the new formalism is consistent.
We can even exchange $\,S\,$ and $\,\ln R\,$
in ordinary quantum mechanics, by taking
$\gamma = \lambda = 1$,
$\kappa = \Lambda = 0$.

It is clear that in this wider
framework, many of the tacit assumptions
of quantum mechanics no longer hold.
For instance, integrability of the probability
density function is only equivalent
to square integrability of the wave function
in certain gauges, so that we are often
outside the usual Hilbert space of quantum
mechanics.

Further details of these results
will be presented elsewhere.

% end of last section 

\section*{Acknowledgments}

I wish to thank
the Alexander von Humboldt
Foundation for generous
support of this work during my
1998-99 sabbatical year in Germany, and
the Arnold Sommerfeld Institute for
Mathematical Physics, Technical University
of Clausthal, for hospitality.

\end{document}